\documentclass[aps,pra,twocolumn,showpacs,superscriptaddress]{revtex4-1}

\usepackage{graphicx}
\usepackage[usenames]{color}
\usepackage{amssymb,amsmath}
\usepackage{leftidx}

\newcommand{\eq}[1]{Eq.~(\ref{#1})}
\newcommand{\fig}[1]{Fig.~\ref{#1}}
\newcommand{\be}[1]{\begin{equation}\label{#1}}
\newcommand{\ee}{\end{equation}}

\begin{document}

\title{The effect of electron-electron correlation on the attoclock experiment }

\author{A. Emmanouilidou}
\affiliation{Department of Physics and Astronomy, University College London, Gower Street, London WC1E 6BT, United Kingdom}
\author{A. Chen}
\affiliation{Department of Physics and Astronomy, University College London, Gower Street, London WC1E 6BT, United Kingdom}

\author{C. Hofmann}
\affiliation{Department of Physics, ETH Zurich, Zurich, Switzerland}
\author{U. Keller}
\affiliation{Department of Physics, ETH Zurich, Zurich, Switzerland}
\author{A. S. Landsman}
\affiliation{Max Planck Institute for the Physics of Complex Systems, Noethnitzer Strasse 38, Dresden, Germany}

\begin{abstract}
We investigate multi-electron effects in strong-field ionization of Helium using a semi-classical model that, unlike other commonly used theoretical approaches, takes into account electron-electron correlation.  
Our approach has an additional advantage of allowing to selectively switch off different contributions from the parent ion (such as the remaining electron or the nuclear charge) and thereby investigate in detail how the final electron angle in the attoclock experiment is influenced by these contributions.  
We find that the bound electron exerts a significant effect on the final electron momenta distribution that can, however, be accounted for by an appropriately selected mean field.  Our results show excellent agreement  with other widely used theoretical models done within a single active electron approximation.  
\end{abstract}

\maketitle

The idea of the attoclock with a clearly defined ``time-zero" for strong field ionization was first introduced in \cite{eckle}, and relies on elliptical polarization of an ultra-short laser pulse to map the final offset angle of the electron momenta distribution to time.  Since the attoclock technique extracts the center of the electron momenta distribution (Fig. 1), the accuracy is limited in principle only by the electron statistics at the detector, hence enabling the reconstruction of electron dynamics in the attosecond ($10^{-18}$ sec) domain.  However, as many reconstruction and experimental calibration procedures in strong field ionization experiments \cite{Nirit,Meckel,Arissian}, the measured delay time  relies on a semiclassical (two-step) model {\it within the single active electron approximation} \cite{tipis,boge,JPhysBTime,optica}.  

Recent experiments \cite{boge,optica} found an additional angular offset, relative to an adiabatic two-step model, which takes account of the Coulomb field of the parent ion within a single active electron approximation (for details on the model see \cite{parabolic,parabolic1,tipis}).  This additional offset was explained as due to tunneling delays, following the attoclock concept, as presented in \cite{eckle,optica,physRep}.  A subsequent work  \cite{JPhysBTime} showed that using an imaginary time method developed by Perelomov, Popov, and Terent'ev (PPT) \cite{b17}, which takes account of non-adiabatic effects but neglects the Coulomb tail inside the potential barrier, gives essentially the same interpretation of the experimental result in \cite{optica} as the above mentioned adiabatic model.   On the numerical front, there remains a discrepancy between time-dependent Schr$\mathrm{\ddot{o}}$dinger Equation (TDSE) simulations performed within the single active electron approximation \cite{praCornelia,Kheifets} and the experimental results in \cite{optica,boge}.  It has therefore been suggested that this discrepancy potentially can be explained by the presence of {\it electron-electron correlations in Helium.}  

Another recent experiment has shown that electron-electron correlations can play an important role in excited Helium atoms when ionized with an ultra-violet photon \cite{heliumPRL}. Most recently, however, excellent agreement was obtained for attosecond photoemission measurements from ground state helium with three different theoretical models, which go from ab-initio theory taking into account both electrons to single-active electron approximation with both the TDSE and a fully analytical model \cite{Keller2015}.  A question remains as to whether such correlation effects are significant in the present experimental range of the attoclock, which involves strong field tunnel ionization  (using a laser pulse with central wavelength in the infrared range) of Helium from a ground state.  Resolving this question with TDSE simulations of a two electron atom in the relevant experimental regime is a formidable task.  
We therefore use a semiclassical two electron model, introduced in \cite{mypaper}, that allows for a consistent description of the process, taking into account the initial bound state and the subsequent propagation in the continuum, but, like other semi-classical models, neglecting a possible tunneling delay.  

To study the effect of electron-electron interaction on the  2-D momentum distribution of the singly ionizing electron in He in an elliptically polarized laser field (i.e. the experimental observable in \cite{boge,optica,landsman2014}), we employ a three-dimensional (3-d) semi-classical model for a three-body Coulomb system (two electrons plus the nucleus). All interactions are fully accounted for including electron-electron correlation. The quasiclassical model we use entails one electron (electron 1) instantaneously tunneling through the field-lowered Coulomb potential with a quantum tunneling rate given by the Ammosov-Delone-Krainov (ADK) formula \cite{parabolic, distribution,b14}. The longitudinal momentum is zero while the transverse one is given by a Gaussian distribution \cite{distribution}. 
We compute the exact exit point for the tunneling electron using parabolic coordinates \cite{parabolic1}. 
The remaining electron (electron 2) is modeled by a microcanonical distribution \cite{microcanonical}.  
For the evolution of the classical trajectories we use the full three-body Hamiltonian in the laser field, that is, 
\begin{equation}
H=\frac{p_{1}^2}{2}+\frac{p_{2}^{2}}{2}-\frac{Z}{r_{1}}-\frac{Z}{r_{2}}+\frac{1}{|\bf{r}_{1}-\bf{r}_{2}|}+(\bf{r}_{1}+\bf{r}_{2})\cdot \bf{E}(t),
\label{eq1}
\end{equation}
with $\mathrm{{\bf E}(t)}$ the laser field. An important difference between our 3-d model and other 3-d semiclassical models \cite{parabolic1,previous} is that our model treats exactly the Coulomb singularity by introducing regularized coordinates \cite{regularized} for the time propagation. Transforming to regularized coordinates removes the Coulomb singularity due to the electron-nucleus interaction. This results in a  faster and more stable numerical integration. The model we currently employ was first described in \cite{mypaper}.

Our goal is to compute the offset angle of the 2-D electron momentum distribution in the plane of polarization as a function of the intensity of the laser field and compare with the experimental results in \cite{boge,optica}. To do so, we model the electric field using a Gaussian pulse, similar to the one employed in the experiments that are of interest for the current work  \cite{boge,experiment1,landsman2014,tipis,optica}:
\begin{eqnarray}
\mathrm{{\bf E}(t)=E_0f(t)(\cos{\omega t} \hat{z}+\epsilon\sin{\omega t}\hat{x})}
\label{eq:dipole}
\end{eqnarray}
with
\begin{eqnarray}
\mathrm{f(t)=\exp\left(-2ln2\left(\frac{t}{t_{FWHM}}\right)^2\right)}
\end{eqnarray}
and $\mathrm{t_{FWHM}=6}$ fs, $\mathrm{\omega=0.05695}$ a.u., with  $\mathrm{E_0}$ the amplitude of the laser field.  We use the ellipticity parameter $\mathrm{\epsilon=0.87}$.   These laser pulse parameters were selected to model the experimental set-up described in \cite{boge,optica}.  The use of the dipole approximation in \eq{eq:dipole}  allows one to neglect the spatial dependence of the laser field as well as the magnetic fields, which would otherwise considerably complicate the dynamics \cite{cohen}, hence allowing for an easier and more intuitive interpretation of classical trajectories \cite{corkum}.
The initial time of propagation is sampled randomly from $\mathrm{\phi_{0}=\omega t_0=[-34,34]}$ a.u., with $\mathrm{\phi_{0}}$ the initial phase of the laser field at the time electron 1 tunnels.

\begin{figure}
\begin{center}
\includegraphics[width=1.0\linewidth]{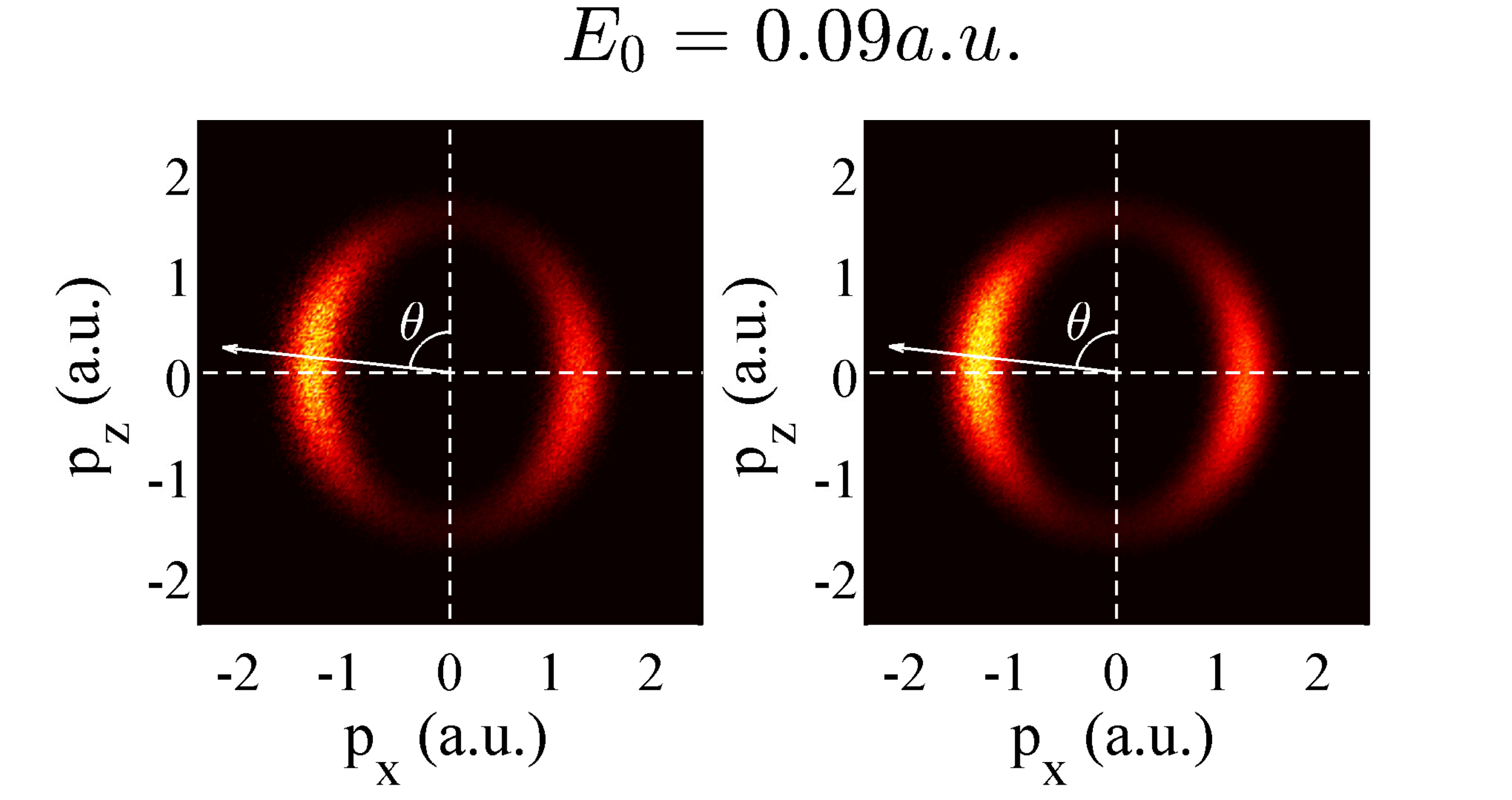}
\caption{The offset angle for the three-body Coulomb system (left) and for the two-body system with an effective charge $\mathrm{Z=1}$ (right). }
\label{fig:offset1}
\end{center}
\end{figure}

We compute the 2-D, x-z, momenta distributions of the ionizing (tunneling) electron, in single ionization events, for different laser field strengths. Specifically, we consider the field strengths   $0.03$, $0.05$, $0.07$, $0.09$, $0.11$ and $0.13$ a.u. In \fig{fig:offset1}, we plot the 2-D momentum distribution for a field strength of $\mathrm{E_{0}=0.09}$ a.u.  We extract the offset angle from these momenta distributions as follows. For each singly ionizing event we register the final momentum of the tunneling electron, which escapes to the continuum. From the final momentum we compute the angle   $\mathrm{\theta=arctan(p_x/p_z)}$ and the magnitude of the momentum $\mathrm{p=\sqrt{p_x^2+p_z^2}}$. The distribution $\mathrm{S(\theta, p)}$ denotes the probability for the singly ionizing electron to escape with angle $\mathrm{\theta}$ and momentum $\mathrm{p}$, respectively. Integrating over the magnitude of the momentum, we then obtain the distribution $\mathrm{f(\theta)}$ that denotes the probability for an electron to singly ionize with angle $\mathrm{\theta}$: 
\begin{eqnarray}
\mathrm{f(\theta)=\int S(\theta,p)pdp}.
\end{eqnarray}
We use a double gaussian function to fit $\mathrm{f(\theta)}$, i.e. the same function as the one used to fit the experimental results in \cite{boge,optica}. It was shown in prior work \cite{hofmann} that, for high ellipticities, 
the radial integration method described above integrates exclusively over the transverse momenta spreads of the electron distribution at the detector.

Using the above described procedure, we extract the maximum, $\mathrm{\theta_{max}}$, of the $\mathrm{f(\theta)}$ distribution, which corresponds to the most probable value of the angle $\mathrm{\theta}$.  We then identify $90^{\circ}-\mathrm{\theta_{max}}$ as the streaking offset angle (see \fig{fig:offset3}).  
$\mathrm{90^{\circ}}$ is how much the momentum at the end of the pulse is offset compared to the laser field at the time of tunneling, assuming no tunneling delays and neglecting the Coulomb interaction with the parent ion.

First, we investigate whether the electron-electron interaction in the  three-body Hamiltonian can be accounted for by using a two-body Coulomb system (one electron plus the nucleus) with an effective nuclear charge equal to $\mathrm{Z_{eff}=1}$. In \fig{fig:offset1}, we plot the 2-D momenta distributions for $\mathrm{E_{0}=0.09}$ a.u. for the three-body and the two-body Coulomb systems. The two distributions are very similar. Indeed, in Table 1 we list the offset angles $90^{\circ}-\mathrm{\theta_{max}}$ for different strengths of the laser field for the three-body and the two-body systems. We find that the two sets of values are very similar. This similarity implies that the electron-electron interaction can be accounted for by a one electron plus a nucleus system with an effective charge. 

\begin{figure}[h]
\begin{center}
\includegraphics[width=1.0\linewidth]{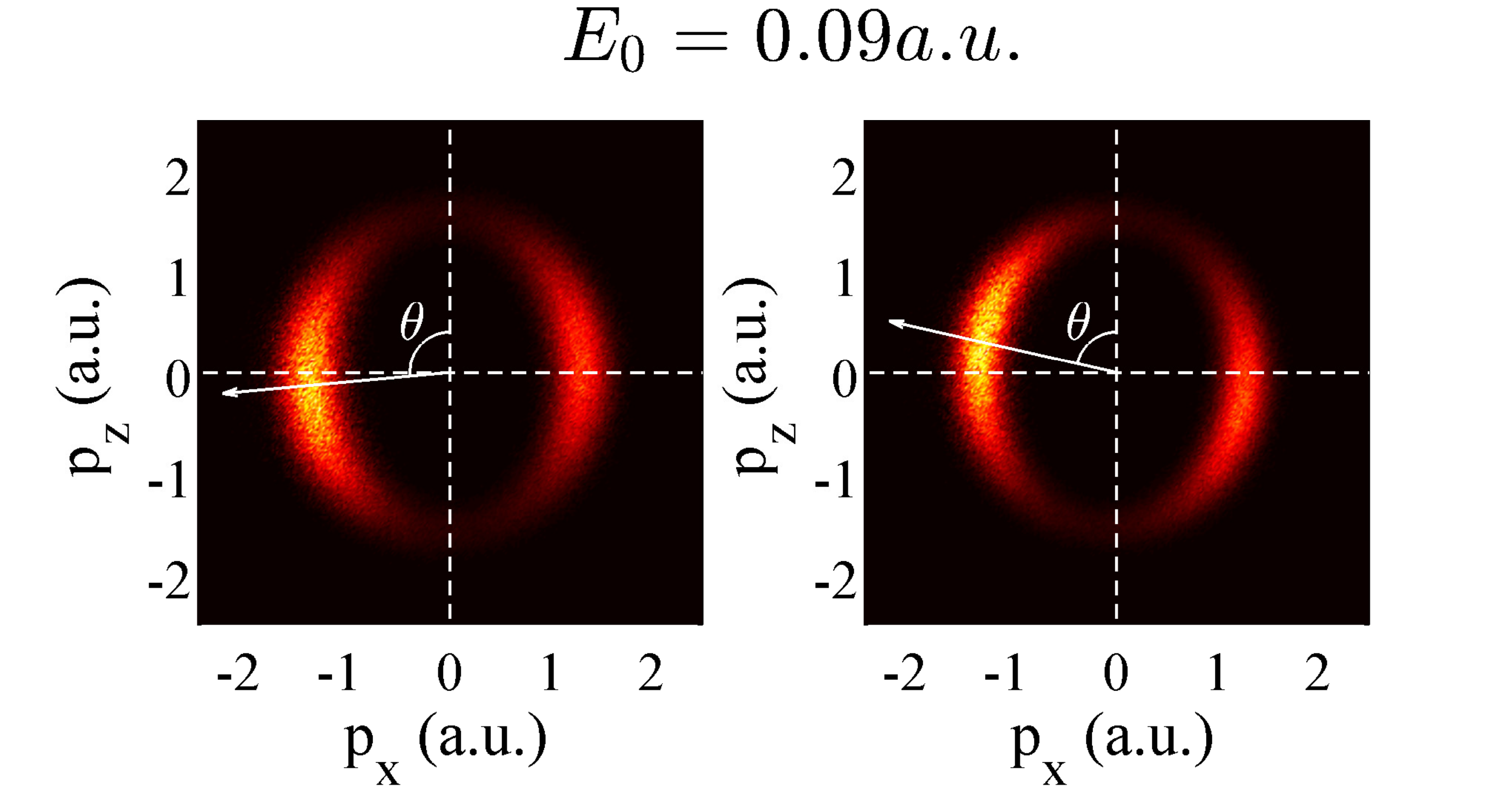}
\caption{ The offset angle for the three-body Coulomb system when the electron-nucleus interaction is switched off for the tunneling electron (left) and for the two-body system with $\mathrm{Z=2}$ (right).  }
\label{fig:offset2}
\end{center}
\end{figure}

\begin{figure}[h]
\begin{center}
\includegraphics[width=1.0\linewidth]{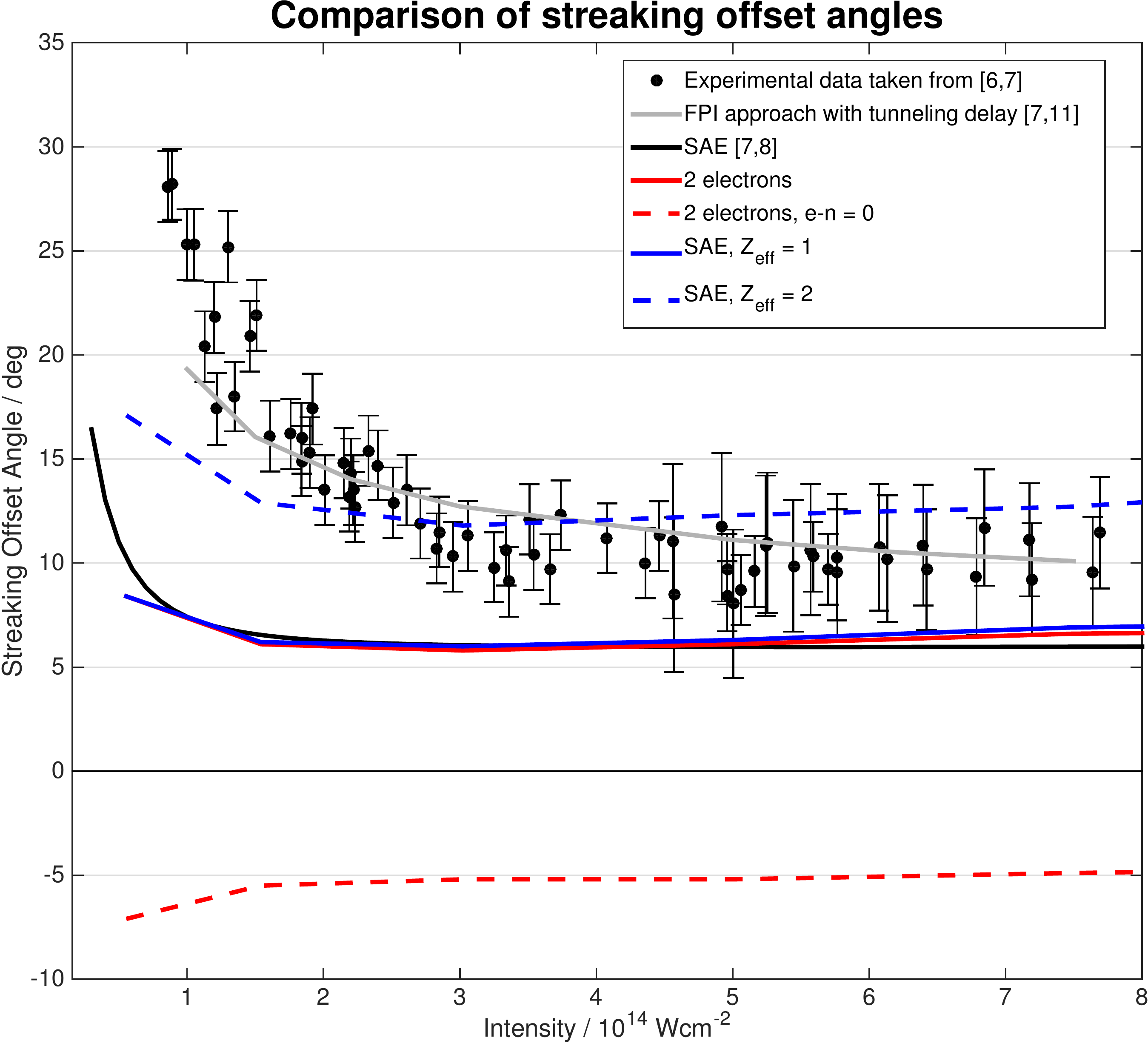}
\caption{Comparison of semiclassical models to experimental results.  All semiclassical models are within a single active electron approximation, except for the solid and dashed red lines.  The colored dashed lines are the result of selectively switching off either the nuclear charge (dashed red line) or  the charge of the remaining electron (dashed blue line).}
\label{fig:offset3}
\end{center}
\end{figure}

Next, we investigate further the effect of electron-electron correlation on the  dynamics of the ionized electron. Specifically,  we compute the offset angle when we switch-off the electron-nucleus interaction following tunneling.  In this case, the offset angle we compute will be due to the effect of the Coulomb repulsion that the tunneling electron experiences from electron 2 that remains bound. In \fig{fig:offset2}, we show for a fixed field strength that  as a result of the electron-electron interaction the offset angle is in the $\mathrm{p_{z}<0}$ \& $\mathrm{p_{x}<0}$ quadrant unlike the offset angles
for the cases of the full three-body system and the two-body system that are in the $\mathrm{p_{z}>0}$ \& $\mathrm{p_{x}<0}$ quadrant. Specifically, as shown over a wide range of electric field strengths in Table II and \fig{fig:offset3}, while for the full three-body and the two-body system the offset angles $90^{\circ}-\mathrm{\theta_{max}}$ are positive, when we switch-off the electron-nucleus interaction on the tunneling electron the offset angle is negative (see Fig. 3, dashed red line). This implies that the Coulomb repulsion between the bound electron  and the tunneling electron causes the latter to escape to the continuum faster than it would in free-space  and even more so if we only account for the effect of the nucleus on the tunneling electron with $\mathrm{Z=2}$, (see Table II and dashed blue curve in \fig{fig:offset3}).
The above implies that the electron-electron interaction (in the mean field sense) does have a significant effect on the dynamics of the tunneling electron while escaping to the continuum. The repulsion from the bound electron causes the tunneling electron to escape faster compared to the case where no electron-electron interaction is present. 

\fig{fig:offset3} shows that the offset angle obtained using a semiclassical model (solid red line) is significantly smaller than the experimental one, particularly for small intensities. Can electron-electron correlation account for this discrepancy? We believe this not to be the case for the following reasons: so far we have shown that, within the framework of our model, the offset angle is the same whether we fully account for two electron effects or in a mean-field sense.  This can be seen in \fig{fig:offset3}, where the two electron (three-body system) results (solid red line) almost coincide with the single active electron approximation (solid blue line), when the effective charge of the remaining parent ion is such that $\mathrm{Z_{eff} = 1}$.  
(In regard to the single active electron model, the effective charge before the ionisation takes place has to be greater than one, since the nuclear charge in a ground state of Helium is only partially shielded by the other electron.) 
In addition, as our results show, electron-electron correlation causes the tunneling electron to escape faster to the continuum resulting in a decreased offset angle.  However, we note that our theoretical results are obtained with a model which does not consider the dynamics inside the barrier. It is only after electron 1 exits the barrier that we propagate in time accounting for all the interactions among the two electrons and the nucleus as well as the interactions with the laser field.  
\fig{fig:offset3} compares our results (solid red and blue lines) with experimental data and with another adiabatic model described in \cite{previous} and referred to as TIPIS in \cite{tipis} (solid black line in \fig{fig:offset3}).  As can be seen, there is good agreement between the semiclassical models over a wide intensity range in which the attoclock experiments in \cite{boge,optica} were performed.  

Investigating electron-electron correlations during the tunneling process itself (while the electron is inside the barrier) would be the next step in ruling out multi-electron effects in the attoclock experiments.  However, such studies are beyond the scope of existing semiclassical models.  
Fully quantum, two-electron, TDSE simulations using a t-SURFF method \cite{Armin1} are currently in the planning stages  \cite{Armin2}.  Such simulations could definitively resolve the importance of electron-electron correlations in Helium, during strong field ionisation from a ground state, by comparing TDSE simulations within a single active electron approximation \cite{praCornelia}, with a full two-electron TDSE solution.  

In conclusion, we note that taking account of electron-electron correlations within a semi-classical model cannot account for the additional angular offset measured in the recent attoclock experiments \cite{boge,optica}.  
In particular, a careful use of the single active electron approximation results in essentially the same angle for the center of the electron momenta distribution as the use of a full two electron model.  One of the advantages of our two-electron model is that it allows for a consistent description of the entire process, both the initial bound state and the subsequent propagation of the ionized electron in the continuum.  On the other hand, the single active electron approximation requires that we use different effective charge for the initial bound atomic state (where it must be such that it reproduces the ionisation potential of Helium) and for the propagation in the continuum, where the total effective charge of the parent ion should be set to one.  
Using the two-electron model also allowed us to investigate the relative contributions of the nucleus and the remaining bound electron to the final offset angle.  


%

 \begin{table} [ht!]  \label{table1}
\begin{center}
\begin{tabular}[t]{|c|c|c|c|c|}
\hline
$\mathrm{E_{0}}$ (a.u.) & Three-body ($^{\circ})$&Two-body, $\mathrm{Z_{eft}=1}$ ($^{\circ})$\\
\hline
0.03&8.4 &8.4\\
\hline
0.05&6.1 &6.2\\
\hline
0.07&5.8 &6.0\\
\hline
0.09&6.1 &6.3\\
\hline
0.11&6.6 &6.9\\
\hline
0.13&6.8&7.2\\
\hline
\end{tabular}
\caption{The offset angle for the three-body Coulomb system (left) and the two body Coulomb system with an effective charge $\mathrm{Z_{eff}=1}$ (right).}
\label{angle_compare1}
\end{center}
\end{table}

%
%

 \begin{table} [ht!]  \label{table1}
\begin{center}
\begin{tabular}[t]{|c|c|c|c|c|}
\hline
$\mathrm{E_{0}}$ (a.u.) &Three-body (a) ($^{\circ}$)&Three-body (b) ($^{\circ}$)& Two-body, $\mathrm{Z=2}$ ($^{\circ})$\\
\hline
0.03&8.4 &-7.1&17.1\\
\hline
0.05&6.1 &-5.5&12.9\\
\hline
0.07&5.8 &-5.2&11.8\\
\hline
0.09&6.1 &-5.2&12.3\\
\hline
0.11&6.6 &-4.9&12.7\\
\hline
0.13&6.8&-4.6&13.9\\
\hline
\end{tabular}
\caption{The offset angle for the three-body Coulomb problem as in  Table 1 (a), for the three-body problem but with the interaction of the tunneling electron with the nucleus switched off (b) and for the two-body problem with $\mathrm{Z=2}$.}\label{angle_compare}
\end{center}
\end{table}

A.E. acknowledges support from EPSRC under Grant No. J0171831  and use of the Legion computational resources at UCL. 
U.K. acknowledges ERC advanced grant ERC-2012-ADG\_20120216 within the seventh framework programme of the European Union and A.S.L. and U.K. acknowledge the support by the Swiss National Science Foundation (SNSF) project grant Nr. 200021\_153432. A.S.L. acknowledges the Max Planck Center of Attosecond Science (MPC-AS).

\pagebreak   
\newpage


\end{document}